\documentclass[12pt]{iopart}

\usepackage{graphicx}
\usepackage{colortbl}
\usepackage{iopams}

\newcommand{\binom}[2]{\left(\begin{array}{c} #1 \\ #2 \end{array}\right)}

\newcommand{\neswarrow}{\mathrel{\mbox{$\nearrow$\llap{$\swarrow$}}}}
\newcommand{\nwsearrow}{\mathrel{\mbox{$\nwarrow$\llap{$\searrow$}}}}

\newcommand{\neswarrowsub}{\mathrel{\mbox{\scriptsize $\nearrow$\llap{$\swarrow$}}}}
\newcommand{\nwsearrowsub}{\mathrel{\mbox{\scriptsize $\nwarrow$\llap{$\searrow$}}}}

\definecolor{mycyan}{rgb}{0,1,1}
\definecolor{mymagenta}{rgb}{1,0,1}
\definecolor{myyellow}{rgb}{1,1,0}

\eqnobysec

\begin{document}

\title[Generalized shifts for polarization components of reflected light beams]{Generalized shifts and weak values for polarization components of reflected light beams}

\author{ J\"org B G\"otte$^{1,2}$ and Mark R Dennis$^2$}
 
\address{$^1$Max-Planck-Institute for the Physics of Complex Systems, Noethnitzer Str. 38, 01187 Dresden, Germany\\
$^2$H H Wills Physics Laboratory, University of Bristol, Tyndall Avenue, Bristol BS8 1TL, UK}
\ead{goette@pks.mpg.de
}

\begin{abstract}
The simple reflection of a light beam of finite transverse extent from a homogenous interface gives rise to a surprisingly large number of subtle shifts and deflections which can be seen as diffractive corrections to the laws of geometrical optics [Goos-H\"anchen shifts] and manifestations of optical spin-orbit coupling [Imbert-Fedorov shifts], related to the spin Hall effect of light. 
We develop a unified linear algebra approach to dielectric reflection which allows for a simple calculation of all of these effects and lends itself to an interpretation of beam shifts as weak values in a classical analogue to a quantum weak measurement. 
We present a systematic study of the shifts for the whole beam and its polarization components, finding symmetries between input and output polarizations and predicting the existence of material independent shifts. 
\end{abstract}



\section{Introduction}\label{sec:intro}

An elementary similarity between ray and wave theories of light is the law of specular reflection: `the angle of incidence equals the angle of reflection'.
This rule, while valid for plane waves, is now understood to be only approximately true for light beams (and, in general, any wave beam) which has finite transverse extent.
As originally discovered by Goos and H\"anchen \cite{GoosHaenchen:AndP436:1947}, under total internal reflection, the centre of a reflected light beam is slightly shifted spatially with respect to the centre of the incident beam in the plane of incidence.
Furthermore, if the incident polarization is neither purely in the plane of incidence nor purely orthogonal to it, the shift has a transverse component too (Imbert-Federov shift) \cite{Imbert:PRD5:1972,Fedorov:DAN105:1955}, which may be interpreted as the spin-orbit coupling of light \cite{LibermanZeldovich:PRA46:1992,BliokhBliokh:PRL96:2006}.
If the reflection is not total, although still dielectric, the reflected beam acquires an angular shift, deflecting the centre slightly upon propagation, which can be both longitudinal and transverse \cite{ChanTamir:OL10:1985,Merano+:NatPhot3:2009,AielloWoerdman:OL33:2008}.
In a closely related effect, different polarizations are reflected slightly differently; for incident linear polarization, different circular components are shifted to opposite sides of the incident plane, although the total beam centre is not transversally shifted, resulting in the `spin Hall effect of light' \cite{HostenKwiat:SCI:2008}. 
The separation of spin components of a beam is, however, not restricted to the transverse direction, and has also been observed within the plane of incidence \cite{Qin+:OE19:2011}.
Basic to all these phenomena is that the magnitude of the shift with respect to a suitable chosen measure is independent of the beam's angular spectrum, provided it is axisymmetric and tightly concentrated about a mean incidence angle.

Unsurprisingly, all of these different shifts are intimately related: angular and spatial, in partial and total reflection, affecting the entire beam and separate polarization components.
Our purpose here is to present a straightforward unified framework for their description and analysis. 
We derive a simple formula for the shift of any polarization component of the reflected beam, and show that the overall centroid of the reflected intensity pattern is a suitably weighted sum of the shifts of orthogonal polarizations of the final beam.
This form is naturally expressed as a complex vector, and writing it in this way reveals a close relationship with the quantum-mechanical notion of weak values and weak measurements of operators \cite{Aharonov+:PRL60:1988,AharonovRohrlich:WVCH:2005}.
In this approach, the effect of an operator on an initial state is not considered overlapped with the same initial state, as with a regular quantum expectation value, but rather with a particular, postselected state which is different from but not orthogonal to the initial state (and neither the initial nor the final state is an eigenstate of the operator).
This has been previously discussed and experimentally measured for the `spin Hall effect of light' \cite{HostenKwiat:SCI:2008}, where a weak measurement was used to show that beam components  undergo shifts that the overall intensity centroid does not experience.

In our case, the initial state preparation corresponds to a uniform initial polarization $\boldsymbol{E},$ a postselecting measurement state from a polarization analyzer $\boldsymbol{F}$ (in general unrelated to $\boldsymbol{E}$), and the operator in question is the reflection matrix, which acts on different plane wave components of the beam differently, since the angle and plane of incidence depend on the direction of the plane wave component. 
For a narrow paraxial beam, this dependence is so small that the polarization couples only weakly to space, so the spatial beam shift is analogous to shift of the pointer.
The weak value of an operator is in general complex, and it has been proved that the imaginary part corresponds to a shift in the conjugate variable \cite{Steinberg:PRA52:1995,Josza:PRA76:2007}.
If the pointer corresponds to the real space representation of the beam, then its conjugate variable is the Fourier space: the imaginary part of the weak value is the angular shift. 
The full extent of the connection between weak values and beam shifts extents even further and deserves its own exposition in a more general framework, which we present in a companion paper \cite{DennisGoette:NJP:2012}.

Here, we consider explicitly the resulting shifts with input and analyzer polarizers ($\boldsymbol{E}$ and $\boldsymbol{F}$) chosen from three natural bases: the transverse electric (TE) and magnetic (TM) (`$+$'), circular (`$\bigcirc$') and $45^{\circ}$ and $135^{\circ}$ linear (`$\times$') bases, and we find a symmetry in the shifts between input and output polarizations common to total and partial reflection.
Surprisingly for a phenomenon which can be used for optical biosensing and nano-probing \cite{YinHesselink:APL89:2006,Rodrigues-Herrera+:PRL104:2010} we also find shifts of polarization components that are material independent.
Furthermore, our unified formula shows that, when measured in the appropriate units, the formula for the angular shift is independent of the shape of the spectrum of the incident beam.
Our approach is self-contained, and our derivation clarifies aspects of previous descriptions of beam shifts.

The physics of beam shifts depends wholly on the Fresnel reflection coefficients $r,$ which are the scalar weightings an electromagnetic plane wave acquires on reflection from a planar reflection surface.
We assume throughout that the reflection is dielectric (i.e.~unit relative magnetic permeability $\mu$ and real relative permittivity $\varepsilon$), and an angle of incidence $\theta >0.$
In general, there are two eigenpolarizations for plane wave reflection, `$s$' corresponding to (electric) polarization perpendicular to the plane of incidence, and `$p$' for (electric) polarization in the incidence plane, with two different Fresnel coefficients \cite{BornWolf:CUP:2003}:
\begin{eqnarray}
   r_s & = \frac{\cos \theta - \rmi \sqrt{\sin^2\theta - n^2}}{\cos \theta + \rmi \sqrt{\sin^2\theta - n^2}} & = \frac{\cos \theta - \sqrt{n^2 - \sin^2\theta}}{\cos \theta + \sqrt{n^2 - \sin^2\theta}}, 
   \label{eq:rs}\\
   r_p & = \frac{n^2 \cos \theta - \rmi \sqrt{\sin^2\theta - n^2}}{n^2 \cos \theta + \rmi \sqrt{\sin^2\theta - n^2}} & = \frac{n^2 \cos \theta - \sqrt{n^2 - \sin^2\theta}}{n^2 \cos \theta + \sqrt{n^2 - \sin^2\theta}}.
   \label{eq:rp}
\end{eqnarray}
Here, $n$ is the relative refractive index (the refractive index of the medium in which the light propagates is in the denominator); that is $n > 1$ for reflection by a denser medium.
The reflection coefficients are each given in two forms, each useful depending on whether $\sin^2\theta - n^2 \lessgtr 0.$
The first assumes that $\sin^2\theta > n^2,$ appropriate for total internal reflection (where $\theta$ is greater than the critical angle $\arcsin n$), in which case the reflection coefficient is complex and unimodular, $|r| = 1.$
In the other case, partial reflection (which also has some transmission, which not considered here), $\sin\theta < n,$ so $r$ is real.
Although the formul\ae are similar, $r_p = 0$ at the Brewster angle $\arctan n,$ whereas $r_s \neq 0$ for dielectric reflection.

The spatial and angular Goos-H\"anchen (GH) shift within the plane of incidence can be intuitively understood from a two-dimensional picture in which $s$ and $p$ polarization can be treated separately. 
Any two-dimensional wave packet or beam consists of a spectrum of plane waves with different incidence angles, so each acquires a different reflection coefficient (for each $s$ and $p$ component of the polarization).
When the beam is tightly concentrated around a common incidence angle $\theta_0,$ all beam shifts are a first-order effect in the spectral width of the beam, and their nature (spatial or angular) depends on how the polarization-weighted sum of $r_s$ and $r_p$ varies over the spectrum: a variation of the imaginary part alone leads to a spatial shift, and real variation, to an angular shift (see below, Fig. \ref{fig:2d}).
Because both spatial and angular Goos-H\"anchen shifts are usually small when compared to a generic measure, they can be seen as corrections to the laws of geometrical optics \cite{Merano+:NatPhot3:2009,AielloWoerdman:OL36:2011}.

\begin{figure}
\begin{center}
\includegraphics[width=\textwidth]{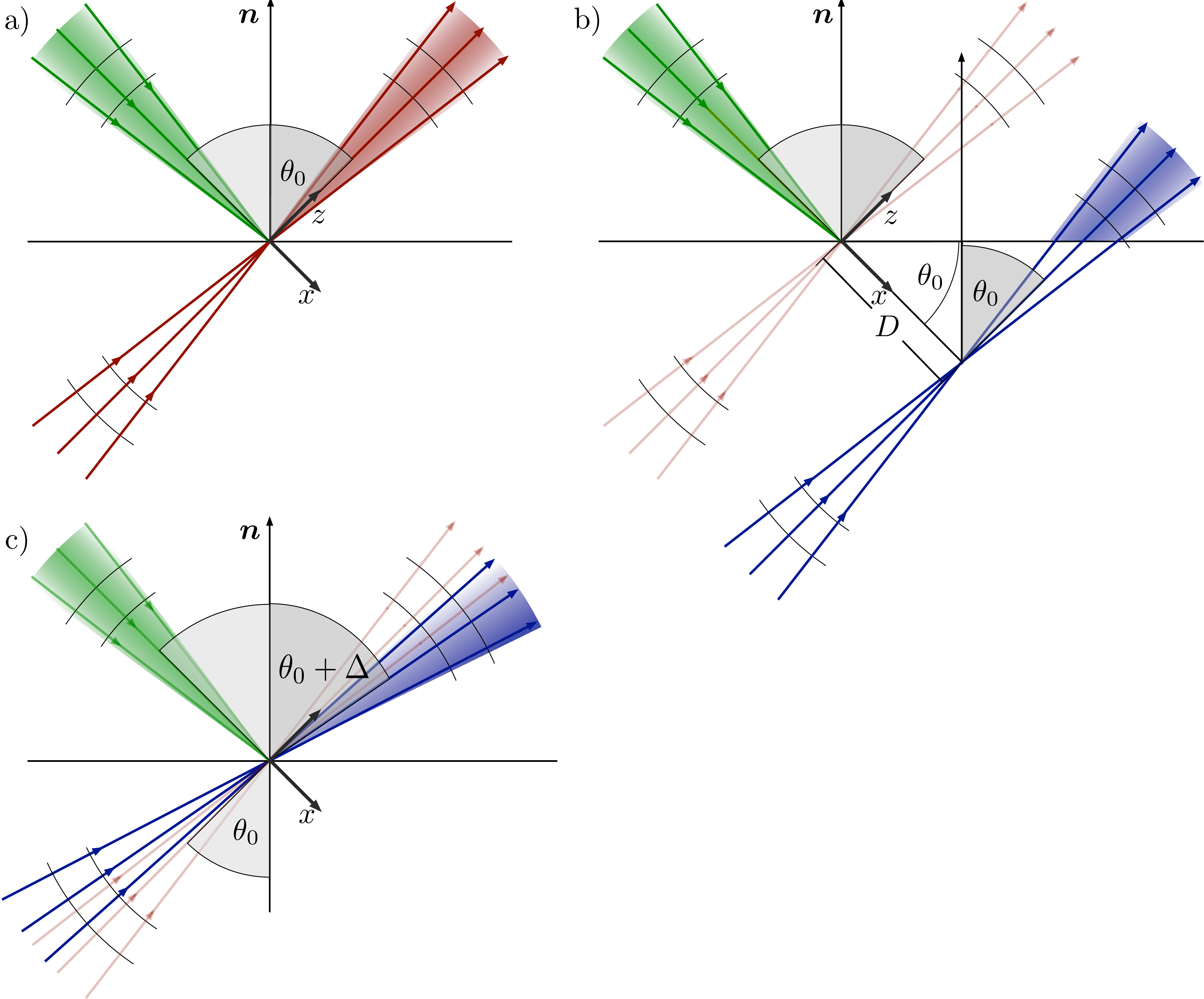}
\caption{\label{fig:2d}
   Schematic illustratiion of `virtual reflection' and spatial and angular Goos-H\"anchen shifts. 
   a) The angular spectrum of the incident light beam (green) is schematically shown as colour gradient. 
   The 'virtual reflected beam' (red) is formed by the geometric reflection with respect to the interface ($x$-axis) of each plane wave in the spectrum. 
   b) For a unimodular reflection coefficient (\ref{eq:rs}), (\ref{eq:rp}) the physically reflected beam appears shifted by distance $D$ parallel to the propagation axis of the virtual reflected beam. 
   c) For a real reflection coefficient the physically reflected beam is deflected by an angle $\Delta$ due to the modulation of the amplitude for each plane wave in the spectrum upon reflection.}
\end{center}
\end{figure}

To understand the Imbert-Federov (IF) shift, which is transverse to the incidence plane, it is necessary to use a vectorial, three-dimensional picture, and it is crucial to distinguish between the global direction of the beam's electrical field and the polarization of the individual plane waves in the angular spectrum.
Owing to transversality, the electrical field for each plane wave in the spectrum has a different direction, and therefore different $s$ and $p$ polarization components with respect to their local plane of incidence.
This is the origin for the `spin-orbit coupling of light' \cite{BliokhBliokh:PRL96:2006} and is responsible for spatial and angular Imbert-Fedorov shifts \cite{Imbert:PRD5:1972}.
In contrast to the in-plane shifts, these transverse shifts do not depend on the change of the reflection coefficient with the angle of incidence, but rather on the difference between the reflection coefficients for the $s$ and $p$ components.

In our calculation, both effects appear in a unified framework by virtue of a general scattering or reflection dyadic $\mathbf{R}$, which projects the electric field of each plane wave component into its local $s$ and $p$ direction, and applies the appropriate Fresnel coefficient to the reflected wave.
The main formul{\ae}  for the total beam and component shifts are given together in Section \ref{sec:formulae}, and are interpreted in terms of weak values of the beam.
These expressions are derived in Section \ref{sec:3dvector}.

We compare the total and component shifts for three naturally incompatible bases of polarizations for the incident and reflected beam in Section \ref{sec:tox}: the $+$ basis given by the transverse electric and magnetic polarizations (i.e.~the eigenpolarizations of 
a plane wave on the beam axis), the $\bigcirc$ basis of circular polarizations, and the $\times$ basis of linear polarizations diagonal to the $+$ basis.
This choice of three bases (represented by three orthogonal directions on the Poincar\'e sphere \cite{BornWolf:CUP:2003}) is clearly special, since each polarization is made up of equal weightings of each vector in the other two bases: for each one of these polarization states, a polarization measurement in a different basis cannot discriminate between this vector and its orthogonal partner.
There are additional symmetries that occur between different polarizations, some of which are specific to whether the reflection is total or partial.


\section{Formul\ae~for total and component beam shifts, and interpretation as 
 values}\label{sec:formulae}

For ease of reference, we present the main formul\ae~for all of the various shifts, including the well-known total beam shift, and the newly derived scalar component shift.
A self-contained calculation for all of them is given in Section \ref{sec:3dvector}.

We think of two different setups, which lead to two different types of shift: the first is the `centre of mass' shift of the entire vector beam, that is without specifying a polarization analyser for the reflected beam. 
The second is the shift of a polarized scalar component of the beam, as filtered out from the non-uniformly polarized reflected beam by use of an analyser. 
The fact that different polarization components undergo different shifts has been called the `spin Hall effect of light' \cite{Onada+:PRL93:2004} when these shifts are transverse, although as we will describe for the general case, both longitudinal and transverse relative shifts between different polarization components are typical.
The total centre of mass shift is, in fact, an average between every pair of orthogonal polarization components, weighted by the intensity of that component (as we will show explicitly later).

We use the coordinates of the reflected beam (as below, Fig.~\ref{fig:vectorcone}), with $x$-, $y$- and $z$- directions corresponding to the axial TM ($p$-direction of the axis), TE (axial $s$-direction), and propagation directions respectively.
(We use TE and TM to avoid confusion with $s$ and $p$, which are defined differently for each plane wave component.)
The polarization of the incident beam in the TM-TE plane is written $\boldsymbol{E}$ and an output polarization component is determined by the analyzer polarization $\boldsymbol{F}$ which in general are complex for elliptic polarization, and given in terms of TE, TM components,
\begin{equation}
   \boldsymbol{E} = (E_m, E_e), \qquad \boldsymbol{F} = (F_m, F_e). \label{eq:pols}
\end{equation}

Our formul\ae~will be written as a complex dimensionless vector $\boldsymbol{\mathcal{D}},$ whose corresponding spatial shift $\boldsymbol{D}$ and angular shift $\boldsymbol{\Delta}$ are given by
\begin{equation}
  \boldsymbol{D} = -\frac{1}{k} \mathrm{Im} \, \boldsymbol{\mathcal{D}}, \qquad 
  \boldsymbol{\Delta} = \langle \delta^2 \rangle \mathrm{Re} \, \boldsymbol{\mathcal{D}},
  \label{eq:reimshift}
\end{equation}
where  $\langle \delta^2 \rangle$ is the second moment of the spectrum in Fourier space, and $k$ is the optical wavenumber.
The $x$-component of these two real vectors, in the TM direction of the beam, is therefore the longitudinal GH shift, and the second, $y$-, component is the transverse IF shift.
The shift formul\ae depend on the beam axis incidence angle $\theta_0,$ the mean reflection coefficients $\overline{r}_p$ and $\overline{r}_s$ evaluated at this angle, and their derivatives $\overline{r}'_p,$ $\overline{r}'_s,$ with respect to $\theta$ at $\theta_0.$ 
The overall beam shift depends on the incident polarization $\boldsymbol{E}_0,$ and the shift of a single polarization component depends on $\boldsymbol{E}$ and $\boldsymbol{F}.$

The expressions for the shifts can be written more compactly with the definition of vectorial quantities relevant to reflection.
The `mean reflection matrix' in the TM-TE basis is 
\begin{equation}
    \overline{\mathbf{R}} = \left(\begin{array}{cc} -\overline{r}_p & 0 \\ 0 & \overline{r}_s \end{array} \right)
    \label{eq:meanrefmat}
\end{equation}
and the first-order term in the small incidence angle expansion of the full matrix depends on two matrices,
\begin{equation}
  \mathbf{R}_\mathrm{GH} = \left(\begin{array}{cc} -\overline{r}_p' & 0 \\ 0 & \overline{r}_s' \end{array} \right), \quad
  \mathbf{R}_\mathrm{IF} = - \cot \theta_0 \left(\begin{array}{cc} 0 & (\overline{r}_s + \overline{r}_p) \\  (\overline{r}_s + \overline{r}_p) & 0 \end{array} \right).
  \label{eq:refmats}
\end{equation}
The diagonal $\mathbf{R}_\mathrm{GH}$ is responsible for the longitudinal shift, whereas the off-diagonal $\mathbf{R}_\mathrm{IF}$ determines the transverse shift.

The shift to the total beam $\boldsymbol{\mathcal{D}}_t$ is simply
\begin{equation}
 \boldsymbol{\mathcal{D}}_t = \left(\frac{\boldsymbol{E}^*\cdot \overline{\mathbf{R}}^\dagger\mathbf{R}_\mathrm{GH}\cdot\boldsymbol{E}}{\boldsymbol{E}^*\cdot \overline{\mathbf{R}}^\dagger\overline{\mathbf{R}}\cdot\boldsymbol{E}}, \frac{\boldsymbol{E}^*\cdot \overline{\mathbf{R}}^\dagger\mathbf{R}_\mathrm{IF}\cdot\boldsymbol{E}}{\boldsymbol{E}^*\cdot \overline{\mathbf{R}}^\dagger\overline{\mathbf{R}}\cdot\boldsymbol{E}}\right),
 \label{eq:finaltotaleq}
\end{equation}
where $\overline{\mathbf{R}}^\dagger$ denotes the conjugate transpose of the mean reflection matrix $\overline{\mathbf{R}}$ and $\boldsymbol{E}^*$ is the complex conjugate of the polarization vector $\boldsymbol{E}$.
In the derivation of this expression in Section \ref{sec:3dvector} we write  $\boldsymbol{\mathcal{D}}_t$ slightly differently, but the expression above is equivalent to (\ref{eq:meanorthref}) using the convention of the 2-component shift vector in which the first and second components refer respectively to the shifts in the longitudinal and transverse direction.

The shift for a polarization component $\boldsymbol{\mathcal{D}}_c$ is given by a similar expression, but now depending on the output polarization $\boldsymbol{F},$
\begin{equation}
  \mathcal{\boldsymbol{\mathcal{D}}}_c= \left( \frac{\boldsymbol{F}^\ast \cdot \mathbf{R}_{\mathrm{GH}} \cdot \boldsymbol{E}}{\boldsymbol{F}^\ast \cdot \overline{\mathbf{R}}  \cdot \boldsymbol{E}}, 
  \frac{\boldsymbol{F}^\ast \cdot \mathbf{R}_{\mathrm{IF}} \cdot \boldsymbol{E}}{\boldsymbol{F}^\ast \cdot \overline{\mathbf{R}}  \cdot \boldsymbol{E}}\right).
  \label{eq:compweak}
\end{equation}
Both shift expressions could be further simplified by writing a 2-vector of $2\times 2$ matrices $(\mathbf{R}_\mathrm{GH},\mathbf{R}_\mathrm{IF}),$ but we will keep the separation explicit in the following.

The form of equation (\ref{eq:compweak}) is very similar to that of a the `weak value' of a quantum-mechanical operator $\hat{A}$ with initial state $|E\rangle$ and final state $|F\rangle$.
The usual definition of the weak value of $\hat{A}$ is \cite{Aharonov+:PRL60:1988}:
\begin{equation}
   A_w = \frac{\langle F | \hat{A} | E \rangle}{\langle F | E \rangle}
   \label{eq:weak}
\end{equation}
The significance of this quantity is that it represents the outcome of the operation $\hat{A}|E\rangle,$ postselected into a new state $|F\rangle$ rather than compared with the initial state $|E\rangle.$
The weak value is of course only defined when $|F\rangle$ is not orthogonal to $|E\rangle$ (otherwise the denominator would blow up the whole expression), but can take on arbitrarily large values when $|F\rangle$ and $|E\rangle$ are almost orthogonal, as the overlap in the numerator is not necessarily particularly small, although the denominator can approach zero arbitrarily.
Hence weak values can be arbitrarily large even when the spectrum of $\hat{A}$ is bounded, and as such are sometimes called `superweak' \cite{BerryShukla:JPA43:2010}.
It is crucial that the interaction of the initial state with the rest of the system is weak, so the final system $\hat{A}|E\rangle$ is largely undisturbed by the action of $\hat{A};$ this weak interaction should involve changing the state of a different, pointer state whose motion indicates the weak value, given by $\mathrm{Re} \, A_w.$
We will discuss this connection briefly below; more details of the connection are given in the companion paper to this one, \cite{DennisGoette:NJP:2012}.

We consider a well collimated beam with a narrow axisymmetric spectrum $\sigma(\delta)$ which is parametrized by a single angle $\delta$, which denotes the angle of every plane wave in the spectrum with the central wave vector or the beam axis (see Fig.~\ref{fig:vectorcone}).  
This beam is in an initial state, namely a wavepacket with a homogeneous initial polarization $\boldsymbol{E},$ represented by a single point in polarization space (the Poincar\'e sphere \cite{BornWolf:CUP:2003}).
After reflection, the beam is strongly localized in polarization space, but is no longer a single point (and, when $\boldsymbol{E}$ is not an eigenpolarization, the state out is not the state in).
Since the initial beam was concentrated in a particular initial direction, the spread of polarization in the reflected beam is small, but postselecting by a particular component $\boldsymbol{F}$ can reveal a larger effect than is present in the overall beam (the total shift).

Furthermore, since the spread in Fourier space is narrow, the beam width is large, much larger than the small (wavelength-scale) shift in the beam's centre.
The configuration space position of the beam thus corresponds to the measurement system, interacting weakly (due to the narrow beam) on reflection with the polarization degrees of freedom: $\mathrm{Re} \, \rmi \boldsymbol{\mathcal{D}}_c$ corresponds to $\mathrm{Re} \, A_w.$

The imaginary part of the weak value (\ref{eq:weak}) is related to a shift in the canonically conjugate variable corresponding to the pointer, i.e.~the direction of the beam (position in Fourier space).
In Ref.~\cite{Josza:PRA76:2007}, it is proved that, under a quantum weak interaction, the weak shift in direction should be proportional to $\mathrm{Im}(A_w) \mathrm{Var}_k,$ where $\mathrm{Var}_k$ is the variance of the wavenumber $k$, which ---  in our case of an axisymmetric spectrum --- is given by $\langle \delta^2 \rangle$, the second moment of the spectrum in the Fourier space.
The case of the coupling of direction and polarization in beam reflection is similar to von Neumann measurement: interaction is instantaneous in propagation distance (i.e.~time), and the weak interaction Hamiltonian (the reflection operator) depends linearly on direction $\boldsymbol{k}.$
Indeed, not only is the angular shift proportional to the imaginary part of $\rmi \boldsymbol{\mathcal{D}}_c,$ but the proportionality factor is precisely the variance of the spectral distribution in Fourier space.

In their experimental verification of the spin Hall effect of light, Hosten and Kwiat \cite{HostenKwiat:SCI:2008} interpreted the very large spatial shift on transmission of an almost vertical analyzer state with a horizontal initial state, on transmission through a dielectric object (we make the reasonable assumption that there are similar transmission counterparts to our equations in Section \ref{sec:formulae}).
In particular, the sign of the shift was seen to depend on the sign of the small perturbation angle from vertical, and of course the shift was not defined when the analyzer was truly vertical.

Of course, the denominators of (\ref{eq:compweak}) and (\ref{eq:weak}) differ, which in the beam shift case depends on the mean reflection matrix $\overline{\mathbf{R}},$ which is (possibly complex) diagonal in the TM-TE basis.
This corresponds to the fact that, unless the initial beam is polarized in the TE or TM direction, the disturbance to the polarization on reflection is not small, and is given by $\overline{\mathbf{R}}.$
The shift is therefore weak with respect to the system (without interaction) evolving into $|\tilde{E} \rangle \equiv\hat{H}|E\rangle,$ corresponding to the polarization of the on-axis plane wave polarization after reflection $\overline{\mathbf{R}}\cdot\boldsymbol{E}.$
With this replacement, we can see that the beam shift truly corresponds to the weak value of the operator we call the `Artmann operator'
\begin{equation}
   (\hat{A}H^{-1})_{\mathrm{weak}} = \frac{\langle F | \hat{A} \hat{H}^{-1} | \tilde{E} \rangle}{\langle F | \tilde{E} \rangle} = \frac{\langle F | \hat{O} | E \rangle}{\langle F | \hat{H} | E \rangle},
   \label{eq:trueweak}
\end{equation}
preselected by the result of the noninteracting evolution operator $\hat{H}$ acting on $|E\rangle,$ and postselected by $|F\rangle.$
Note that the notation here differs slightly from \cite{DennisGoette:NJP:2012}.

Superweak values are therefore guaranteed when the postselection analyzer $\boldsymbol{F}$ is almost orthogonal to $\overline{\mathbf{R}}\cdot\boldsymbol{E},$ rather than almost orthogonal to $\boldsymbol{E}$ itself.
For the choice in \cite{HostenKwiat:SCI:2008}, $\boldsymbol{E}_0$ is an eigenstate of $\overline{\mathbf{R}},$ so in this case only should $\boldsymbol{F}$ be almost perpendicular to $\boldsymbol{E}.$

A different, arbitrarily large shift is possible, distinct from the aforementioned `superweak shift': the mean reflection matrix $\overline{\mathbf{R}}$ itself may be close to singular, which occurs when $\theta_0$ is close to the Brewster angle (at which $\overline{r}_p = 0$).
To take advantage of this large shift, either the analyzer or the numerator should be wholly longitudinal, say the analyzer $\boldsymbol{F}.$
In this case, $\boldsymbol{F}^*\cdot\overline{\mathbf{R}}\cdot \boldsymbol{E} = -\overline{r}_p E_m,$ which is arbitrarily small regardless of the initial polarization.
Of course, this discussion is only valid in the first order in $\delta$ approximation we have developed, and more sophisticated mathematics has been employed to describe reflection at Brewster incidence \cite{AielloWoerdman:arxiv:2007,Aiello+:OL34:2009a,GoetteDennis:OL:2012}.

Therefore, given the presence of the mean reflection operator in the denominator of (\ref{eq:compweak}), there should be a finite but nonzero shift between any pair of orthogonal polarizations.
After the derivation of the shift formul\ae, given in the next section, we will discuss the spatial and angular shifts that occur between states in the $+,$ $\bigcirc$ and $\times$ bases, for both total and partial reflection.
However, first we give a derivation of the beam shift formul\ae.
Readers uninterested in these derivations, but are interested in their consequences, may skip to Section \ref{sec:tox}.


\section{Calculation of beam shift formul\ae}\label{sec:3dvector}

Our approach to the calculation of the shift formul\ae given in the previous section is based on the introduction of a `virtual reflected beam', which is in reflected beam coordinates, constructed by `mirroring' every plane in the angular spectrum as in geometrical optics without taking the reflection coefficient into account.
Effectively, the virtual beam is the reflection of the incident beam with $r_s = 1, r_p = 1$ for all directions, but otherwise the virtual beam incorporates the direction change of each plane wave component on reflection.
It is therefore the virtual beam which is depicted in Fig.~\ref{fig:2d} (a).
The real reflected beam follows from applying the reflection coefficients (\ref{eq:rs}), (\ref{eq:rp}) to each plane wave component of the virtual beam separately.

We find the virtual beam a useful conceptual object in considering beam shifts, particularly the angular shift.
The origin of the angular shift comes from the fact that different plane wave components acquire different amplitude weightings, as the real reflection coefficient depends on angle.
In Fourier space, the shift comes from a real gradient applied to the spectrum, as in Fig.~\ref{fig:2d}(c), rather than a translation of the overall direction (which would be closer to the spatial shift).
The virtual beam is the same, independent of the physics of the different physical regimes considered (spatial or angular, total or partial reflection).
We note that the virtual beam utilized here is slightly different from the one we describe in \cite{DennisGoette:NJP:2012} -- there, the virtual beam has a different overall polarization from that considered here.

We therefore work in beam coordinates defined above, with $x$, $y$ and $z$ the TM, TE and propagation directions.
In coordinates based on the dielectric interface (implicit in Fig.~\ref{fig:2d}), the reflection angle of the beam is $\theta_0,$ which therefore corresponds to the $z$-direction.
In spherical polar coordinates based around the propagation direction, we define an azimuth angle $\alpha$ and colatitude $\delta,$ which is zero on the beam axis.
Each plane wave making up the beam has a wavevector of magnitude $k,$ and is parametrized by $\alpha$ and $\delta,$ 
\begin{equation}
  \boldsymbol{k}(\delta,\alpha) = k \boldsymbol{\kappa}(\delta,\alpha) = k  (\cos\alpha \sin \delta, \sin\alpha \sin \delta,\cos\delta).
  \label{eq:cone}
\end{equation}
The spectrum, giving the real amplitude weighting of the plane waves, is denoted $\sigma(\delta),$ which is concentrated around $\delta = 0,$ that is, the vector $(0,0,1)$ along the propagation axis.
We also assume that the spectrum is independent of $\alpha$ (i.e.~axisymmetric), and is square-integrable, but apart from this (and the concentration around $\delta = 0$), the spectrum is otherwise completely general. 
By transversality, the initial polarization of the axial $(0,0,k)$ component beam must be perpendicular, with components $E_m, E_s$ in the TM, TE directions (and given by the 2-vector $\boldsymbol{E}$ in this plane).
Since the incident beam has a constant overall polarization given by the constants $E_m, E_e,$ the polarization of each plane wave component must have these components in the TM, TE direction.

\begin{figure}
\begin{center}
\includegraphics[width=0.8\textwidth]{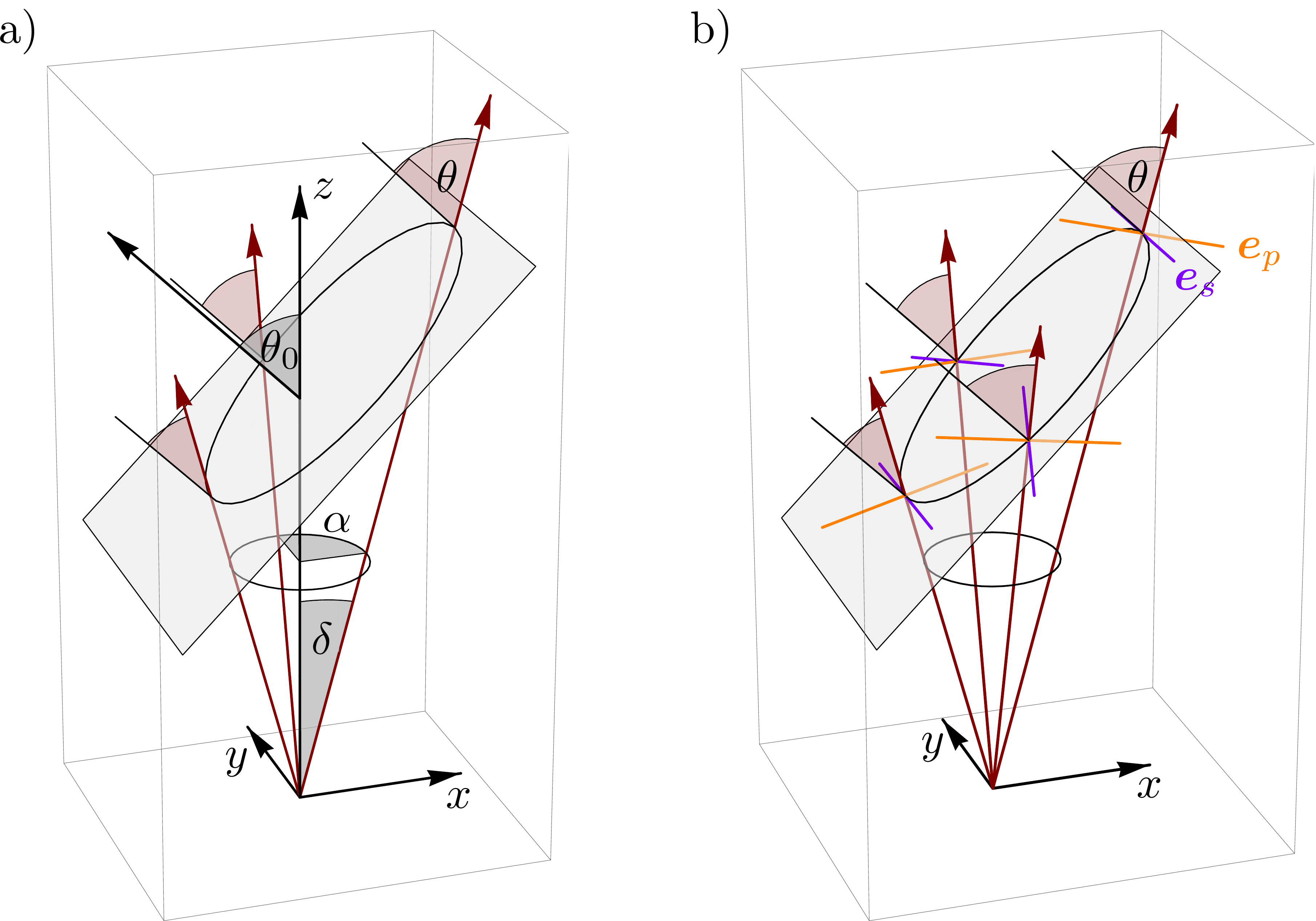}
\caption{\label{fig:vectorcone}Schematic explanation of the beam coordinate system $x,y$ and $z$, where $z$ points along the propagation direction of the beam, $x$ lies in the global plane of incidence and $y$ is transverse to both $x$ and $z$. a) shows the cone geometry used to parametrize the angular spectrum by $\delta$ and $\alpha$ and illustrates that the local angle of incidence $\theta$ varies around the angular spectrum. b) The local planes of incidence are not coplanar with the $x,z$ plane and this is why the decomposition of the polarisation into the local $s$ and $p$ directions varies around the spectrum.}
\end{center}
\end{figure}

The polarization of the incident beam, and hence the virtual beam, is uniform polarized. 
However, since the plane of incidence varies around the beam azimuth $\alpha,$ the precise resolution into each plane wave's particular $s$ and $p$ direction varies around the beam (\Fref{fig:vectorcone}). 
A direct consequence of this variation is that the reflected beam will no longer be uniformly polarized, which is why choosing different settings for the polarization analyzer $\boldsymbol{F}$ can lead to very different beam shifts.
The shifts themselves only depend on terms to first order in $\delta$ in the reflection coefficient and local incidence direction; higher-order terms have a smaller effect which we ignore.

One such ignorable effect comes from the fact that each plane wave acquires a component proportional to $\tan \delta$ in the $(0,0,1)$ direction, which couples into the transverse plane by projection into the local $s,p$ directions: this coupling is also of order $\delta,$ so this extra spin-orbit effect plays no significant role in the transverse shifts, though it affects the axial polarization \cite{Aiello+:PRL103:2009}.
The whole analysis can therefore be restricted to the two transverse dimensions alone.
The virtual reflected beam in real space, $\boldsymbol{E}_v,$ is therefore given by
\begin{equation}
  \boldsymbol{E}_v(\boldsymbol{r}) = \int_0\rmd\delta \int_{-\pi}^{\pi}\rmd\alpha\, \sigma(\delta) \sin\delta  \exp[\rmi k \sin\delta(x \cos\alpha+y \sin \alpha)] \boldsymbol{E},
  \label{eq:virtualvector}
\end{equation}
where $x,y$ are TM, TE beam coordinates, and we ignore $z$-dependence. 
We are not concerned with the upper limit in the $\delta$ integral, as it is cut off by the narrow spectrum $\sigma(\delta).$

In beam coordinates, the reflection plane is now oblique, as depicted for a cone of wavevectors with constant $\delta$ in Fig.~\ref{fig:vectorcone}.
The intersection of the $k$-space cone and the plane is thus a conic section, and the precise angle of incidence of each vector on the cone varies with $\alpha.$
If the reflection angle of the beam axis is $\theta_0,$ the normal vector to the reflection plane in beam coordinates is
\begin{equation}
  \boldsymbol{n} = (-\sin\theta_0,0,\cos\theta_0)
  \label{eq:normaldef}
\end{equation}
so the beam coordinate $y$-direction is perpendicular to the plane of incidence.
The angle of incidence of each vector on the cone is therefore given by $\cos \theta = \boldsymbol{n} \cdot \boldsymbol{\kappa}$ or
\begin{equation}
   \cos\theta(\theta_0, \delta, \alpha) = \cos \delta \cos \theta_0 - \cos \alpha \sin \delta \sin \theta_0.
   \label{eq:localcos}
\end{equation}

Whatever the overall incoming polarization $\boldsymbol{E},$ it is resolved into the local $s, p$ directions for each component wavevector labelled by $\alpha, \delta,$ each of which also acquires the appropriate reflection coefficient $r_s, r_p$ from (\ref{eq:rs}), (\ref{eq:rp}), with its local angle of incidence provided by (\ref{eq:localcos}).
As we are only interested in effects linear in $\delta,$ these coefficients effectively become, for $j = s,p,$
\begin{equation}
  r_j \approx \overline{r}_j + \delta \cos \alpha \: \overline{r}_j \approx \overline{r}_j \exp(\delta \cos \alpha \: \overline{r}'_j/\overline{r}_j).
  \label{eq:rjapprox}
\end{equation}
The second approximation is only valid if $\overline{r}_j \neq 0$, i.e.~not at the Brewster angle.

For a given wavevector $\boldsymbol{\kappa}(\delta,\alpha),$ the $s$ polarization perpendicular to the local plane of incidence is given by the direction of the vector product with the normal direction $\boldsymbol{n},$ namely
\begin{equation}
  \boldsymbol{e}_s(\delta,\alpha) = \frac{\boldsymbol{n}\times\boldsymbol{\kappa}(\delta,\alpha)}{|\boldsymbol{n}\times\boldsymbol{\kappa}(\delta,\alpha)|}.
  \label{eq:esdef}
\end{equation}
The $p$ polarization vector is the cross product of $\boldsymbol{e}_s$ with $\boldsymbol{\kappa},$
\begin{equation}
  \boldsymbol{e}_p(\delta,\alpha) = \boldsymbol{e}_s(\delta,\alpha)\times\boldsymbol{\kappa}(\delta,\alpha).
  \label{eq:epdef}
\end{equation}
These vectors give rise to projection matrices in the TM,TE plane, which are then also approximated up to order $\delta,$
\begin{eqnarray}
  \mathbf{P}_s & = (\boldsymbol{e}_s \otimes \boldsymbol{e}_s) & \approx \left(\begin{array}{cc} 0 & -\delta \cot\theta_0\sin\alpha \\ -\delta \cot\theta_0\sin\alpha & 1 \end{array}\right), \label{eq:ps} \\
  \mathbf{P}_p & = (\boldsymbol{e}_p \otimes \boldsymbol{e}_p) & \approx \left(\begin{array}{cc} 1 & \delta \cot\theta_0\sin\alpha \\ \delta \cot\theta_0\sin\alpha & 0 \end{array}\right). \label{eq:pp}
\end{eqnarray}
Clearly, the variation of order $\delta$ in these projection matrices is proportional to $\sin\alpha,$ as can be seen in Fig.~\ref{fig:vectorcone}.
When $\alpha = 0,$ or $\pi,$ the local plane of incidence agrees with the plane of incidence of the beam axis, so there is no additional polarization effect.
This dependence on $\sin\alpha$ has been interpreted as a spin-orbit effect \cite{BliokhBliokh:PRL96:2006}, as it implicitly involves considering the local polarization of the beam (spin) around the beam azimuth (orbit).

The total transverse reflection dyadic, then, is the sum of the projection matrices weighted by the appropriate reflection coefficient. 
Before we do so, however, we point out a subtlety made explicit by the introduction of the virtual beam: since the virtual reflected beam comes from reflection, it involves a change in the handedness of the coordinate system (right-handed circular light is reflected as left-handed, and vice versa). 
Nevertheless, it is natural to make the coordinate systems attached to the incident, virtual and real reflected beams right-handed, so for every plane wave, the triple of wavevector $\boldsymbol{k}$, electrical field, and magnetic field has a positive scalar triple product.
To account for this discrepancy we choose introduce an additional sign change in the reflection coefficient $r_p$, which leaves the transverse TE ($y$) direction unchanged.
With this sign convention the total reflection dyadic is
\begin{eqnarray}
  \fl \mathbf{R}  = r_s \mathbf{P}_s - r_p \mathbf{P}_p  & \approx & \left(\begin{array}{cc} -\overline{r}_p & 0 \\ 0 & \overline{r}_s \end{array}\right) + \delta \left(\begin{array}{cc} -\overline{r}'_p \cos\alpha & -(\overline{r}_s + \overline{r}_p)\cot \theta_0 \sin\alpha \\ -(\overline{r}_s + \overline{r}_p)\cot\theta_0\sin\alpha & \overline{r}'_s \cos\alpha \end{array}\right) \nonumber \\
  & = & \overline{\mathbf{R}} + \delta \cos \alpha \mathbf{R}_{\mathrm{GH}} + \delta \sin \alpha \mathbf{R}_{\mathrm{IF}},
  \label{eq:dyadic}
\end{eqnarray}
plus terms of order $\delta^2$ and above.
The second line provides the origin of the matrices introduced in (\ref{eq:meanrefmat}), (\ref{eq:refmats}) above.

We are now in a position to compute the beam shifts given in Section \ref{sec:formulae} above.
The total reflected beam $\boldsymbol{E}_r(\boldsymbol{r})$ is constructed by putting the first-order polarization dyadic (\ref{eq:dyadic}) into the virtual beam (\ref{eq:virtualvector}),
\begin{equation}
  \fl \boldsymbol{E}_r(\boldsymbol{r}) = \int_0\rmd\delta \int_{-\pi}^{\pi}\rmd\alpha\, \sin\delta \sigma(\delta) \exp[\rmi k \sin\delta(x \cos\alpha+y \sin \alpha)] \mathbf{R} \cdot\boldsymbol{E}.
  \label{eq:vectorref}
\end{equation}
A scalar component of the reflected field is selected by projecting this beam onto the direction of a polarization analyzer $\boldsymbol{F}$ (whose components are constant). 
The reflected scalar component is thus given by
\begin{equation}
  \fl \boldsymbol{F}^\ast \cdot \boldsymbol{E}_r(\boldsymbol{r}) = \int_0\rmd\delta \int_{-\pi}^{\pi}\rmd\alpha\, \sin\delta \sigma(\delta) \exp[\rmi k \sin\delta(x \cos\alpha+y \sin \alpha)]  \boldsymbol{F}^\ast \cdot \mathbf{R} \cdot \boldsymbol{E}.
  \label{eq:scalarref}
\end{equation}
The scalar $\boldsymbol{F}^\ast \cdot \mathbf{R} \cdot \boldsymbol{E}$ thus determines the nature of the general shift and it is not necessary to calculate the integral in Eq.~(\ref{eq:scalarref}) explicitly.

The scalar component of the reflected beam (\ref{eq:scalarref}) only depends on an angle-dependent reflection scalar
\begin{eqnarray}
  \fl R = \boldsymbol{F}^\ast \cdot \mathbf{R} \cdot \boldsymbol{E}
  & \approx & \overline{R} + \delta (\cos \alpha \, \boldsymbol{F}^\ast \cdot \mathbf{R}_{\mathrm{GH}} \cdot \boldsymbol{E} + \sin \alpha \,\boldsymbol{F}^\ast \cdot \mathbf{R}_{\mathrm{IF}} \cdot \boldsymbol{E}), \nonumber \\
  & = & \overline{R} \left[ 1 + \delta \boldsymbol{\mathcal{D}}_c \cdot(\cos \alpha, \sin\alpha)\right],
\end{eqnarray}
where $\boldsymbol{\mathcal{D}}_c$ is the complex shift vector dependent on $\boldsymbol{F}, \boldsymbol{E},$ defined in (\ref{eq:compweak}), and $\overline{R} \equiv \boldsymbol{F}^\ast \cdot \overline{\mathbf{R}} \cdot \boldsymbol{E},$ the mean  reflection coefficient for the scalar component given by $\boldsymbol{F}.$
Using the same reasoning as in the expansion of the reflection coefficient in Eq.~(\ref{eq:rjapprox}), $R$ may be expressed as an exponential,
\begin{equation}
  R \approx \left(\boldsymbol{F}^\ast \cdot \overline{\mathbf{R}} \cdot \boldsymbol{E}  \right) \exp \left[ \delta \boldsymbol{\mathcal{D}}_c\cdot(\cos \alpha, \sin\alpha) \right].
  \label{eq:scalarcompexp}
\end{equation}
The trigonometric factors $\cos \alpha$  and $\sin \alpha$ therefore determine which terms in the exponential are responsible for the respective shifts in the transverse and longitudinal directions, because inside the exponential in Eq.~(\ref{eq:scalarref}) $\cos \alpha$ couples to $x$ whereas $\sin \alpha$ couples to $y.$
Alternatively, the $x$-component of $\boldsymbol{\mathcal{D}}_c$ contains all those terms in the exponent of $R$ in Eq. (\ref{eq:scalarcompexp}) which have $\cos \alpha$ as a common factor, while the $y$ component collects all the terms which multiply $\sin \alpha$.

To see that this complex vector really gives rise to both angular and spatial shifts it is easiest to split $\boldsymbol{\mathcal{D}}_c$ into real and imaginary parts explicitly,
\begin{equation}
  \fl R \approx \overline{R} \left[ 1 + \delta \mathrm{Re} (\boldsymbol{\mathcal{D}}_c)\cdot(\cos \alpha, \sin \alpha) \right] \exp \left[ \rmi \delta \mathrm{Im} (\boldsymbol{\mathcal{D}}_c)\cdot(\cos \alpha, \sin \alpha) \right],
\end{equation}  
with the imaginary part in the exponent.
Substituting this expression for $\boldsymbol{F}^* \cdot \mathbf{R}\cdot\boldsymbol{E}$ into the expression (\ref{eq:scalarref}) of the reflected beam and identifying $\delta \approx \sin \delta$ in the exponential, we construe the square brackets containing the real part of $\boldsymbol{\mathcal{D}}_c$ as a direction-dependent modification of the spectrum $\sigma(\delta),$ and the exponential with the imaginary part as a shift of the spatial variables within the Fourier kernel. 
Setting $\boldsymbol{D}_c = (D_{c,x}, D_{c,y})$ (see Eq. (\ref{eq:reimshift})) in the transverse plane gives 
\begin{eqnarray}
\boldsymbol{F}^\ast \cdot \boldsymbol{E}_r(\boldsymbol{r}) & = \overline{R} \int_0\rmd\delta \int_{-\pi}^{\pi}\rmd\alpha\, \sin\delta \sigma(\delta) \left[ 1 + \delta \mathrm{Re} (\boldsymbol{\mathcal{D}}_c).(\cos \alpha, \sin \alpha) \right] \nonumber \\
& \quad \times \exp[ \rmi k \sin\delta \, (\boldsymbol{r}- \boldsymbol{D}_c) \cdot (\cos\alpha,\sin\alpha)]. 
   \label{eq:shiftedcomponent}
\end{eqnarray}
The effect of the imaginary part of $\boldsymbol{\mathcal{D}}_c$ has therefore been to translate the transverse spatial coordinates of the beam by $\boldsymbol{D}_c,$ given in (\ref{eq:reimshift}) for the particular choice of component.
The change in the spectrum does not affect the spatial variables to leading order in $\delta.$

The angular shift is calculated as the expectation value of the spectrum of the reflected component as a distribution in Fourier space.
This distribution is the modulus squared of the spectrum of (\ref{eq:shiftedcomponent}), which is equal to $|\sigma(\delta)|^2,$ unaffected by the unimodular spatial shift, which only affects the spectrum's phase.
We therefore calculate the centroid of an angle vector $(\delta_x,\delta_y) = \delta (\cos \alpha, \sin \alpha)$, whose components represent the longitudinal and transverse coordinates in Fourier space about the mean propagation direction $(0,0),$ which is also the centroid of the virtual reflected beam.
The correctly-normalized expectation value in Fourier space is therefore
\begin{equation}
  \fl \langle \delta_x, \delta_y \rangle = \frac{\int_0\rmd\delta \int_{-\pi}^{\pi}\rmd\alpha\, |\sigma(\delta)|^2 \sin\delta  \left[1+ \delta \mathrm{Re} (\boldsymbol{\mathcal{D}}_c)\cdot(\cos \alpha, \sin \alpha) \right]^2  \delta (\cos \alpha, \sin \alpha)}{\int_0\rmd\delta \int_{-\pi}^{\pi}\rmd\alpha\, |\sigma(\delta)|^2 \sin\delta \, \left[1 + \delta \mathrm{Re} (\boldsymbol{\mathcal{D}}_c)\cdot(\cos \alpha, \sin \alpha)\right]^2}.
\end{equation} 
For the evaluation of this integral the term $\left[1 + \delta \mathrm{Re} (\boldsymbol{\mathcal{D}}_c)\cdot(\cos \alpha, \sin \alpha)\right]^2$ is crucial, and 
keeping only the the two lowest orders in $\delta,$ it can be approximated to $1 + 2 \delta \mathrm{Re} (\boldsymbol{\mathcal{D}}_c)\cdot(\cos \alpha, \sin \alpha)$ both in the numerator and denominator. 
Utilising the fact that $\cos$ and $\sin$ are even and odd functions it is straight forward to evaluate the integral which gives the final centroid as
\begin{equation}
   \langle \delta_x, \delta_y \rangle = \boldsymbol{\Delta}_c,
 \end{equation}
defined for the beam's component by (\ref{eq:reimshift}), with the second spectral moment $\langle \delta^2 \rangle$ given by
\begin{equation}
   \langle \delta^2 \rangle \equiv \frac{\int_0\rmd\delta |\sigma(\delta)|^2 \delta^2 \sin\delta}{\int_0\rmd\delta |\sigma(\delta)|^2 \sin\delta }
\end{equation}
As indicated above, using the second spectral moment as the appropriate unit for the angular shift makes the the angular shift independent of the spectrum when the shift is measured with respect to $\langle \delta^2 \rangle.$
This measure is different for a two dimensional wave packet, accounting for the factor $2$ in the formula for angular shift of a quasi-scalar  `sheet' of light \cite{Merano+:OL21:2010}.

With the shifts now defined for any polarization component of the output beam, it is possible to rederive the shift of the overall intensity pattern of the vector reflected beam $\boldsymbol{E}_r.$
The complex shift vector $\boldsymbol{\mathcal{D}}$ for this total `centre of mass' shift is a sum of the shift vectors for two orthogonal polarization components as analyser settings, weighted by intensity. 
If one such analyzer polarization is given by $\boldsymbol{F} = (F_m, F_e),$ its orthogonal partner can be denoted $\boldsymbol{F}_{\bot} = (F_e^*,-F_m^*),$ so $\boldsymbol{F}^\ast \cdot \boldsymbol{F}_\bot = 0.$
The mean reflection coefficient $\overline{R} = \boldsymbol{F}^*\cdot\overline{\mathbf{R}}\cdot\boldsymbol{E}$ is, to leading order, the amplitude of the scalar component of the reflected beam corresponding to $\boldsymbol{F},$ and 
\begin{equation} \label{eq:meanorthref}
   \overline{R}_{\bot} \equiv  \boldsymbol{F}_\bot^*\cdot\overline{\mathbf{R}}\cdot\boldsymbol{E} = -F_m E_e^{\vphantom{\ast}} \overline{r}_s - F_e E_m^{\vphantom{\ast}} \overline{r}_p
\end{equation}
the amplitude for the orthogonal analyzer component $\boldsymbol{F}_{\bot},$ which has a complex shift vector $\boldsymbol{\mathcal{D}}_{c,\bot}$ given by (\ref{eq:compweak}) with $\boldsymbol{F}_{\bot}$ instead of $\boldsymbol{F}.$
The sum of these two complex shift vectors, weighted by their respective intensities $|\overline{R}|^2, |\overline{R}^{\bot}|^2$ is therefore
the total reflected beam is given by
\begin{equation} \label{eq:totaleq}
  \boldsymbol{\mathcal{D}} = \frac{|\overline{R}|^2 \boldsymbol{\mathcal{D}}_c + |\overline{R}_\bot|^2 \boldsymbol{\mathcal{D}}_{c,\bot}}{|\overline{R}|^2 + |\overline{R}_\perp|^2},
\end{equation}
which gives the $\boldsymbol{F}, \boldsymbol{F}_\bot$-independent form (\ref{eq:finaltotaleq}) after straightforward but tedious algebraic manipulation.
We have therefore given an alternative definition of the total shift vector of a vector beam as a weighted sum of its more fundamental scalar components.


\section{Centre of mass and scalar component shifts in the `$+\bigcirc\times$' bases}\label{sec:tox}

In this section we illustrate the variety of beam shifts present in the equations (\ref{eq:compweak}) and (\ref{eq:finaltotaleq}) for different choices of incident and analyser polarization. We choose three bases of orthogonal polarizations as settings for the incident polarisation $\boldsymbol{E}$ and the analyser polarisation $\boldsymbol{F}$ and consider all possible combinations of pre- and postselection for these bases. The three bases are TM and TE polarisation (`$+$'), right and left circular polarization (`$\bigcirc$') and the two diagonal polarizations of 45$^\circ$ and 135$^\circ$ degrees (`$\times$'). On using the same notation for the components of the polarisation above, the `$+\bigcirc\times$' bases are explicitly given by the following polarization vectors:
\begin{eqnarray}
  \fl \boldsymbol{E}^\leftrightarrow = \boldsymbol{F}^\leftrightarrow = \binom{1}{0}, &\: \boldsymbol{E}^\circlearrowleft = \boldsymbol{F}^\circlearrowleft = \frac{1}{\sqrt{2}}\binom{1}{\mathrm{i}}, &\: \boldsymbol{E}^{\neswarrowsub}= \boldsymbol{F}^{\neswarrowsub}= \frac{1}{\sqrt{2}}\binom{1}{\mathrm{1}} \\
    \fl \boldsymbol{E}^\updownarrow = \boldsymbol{F}^\updownarrow = \binom{0}{1}, &\: \boldsymbol{E}^\circlearrowright = \boldsymbol{F}^\circlearrowright = \frac{1}{\sqrt{2}}\binom{1}{-\mathrm{i}}, &\: \boldsymbol{E}^{\nwsearrowsub} = \boldsymbol{F}^{\nwsearrowsub} = \frac{1}{\sqrt{2}}\binom{1}{\mathrm{-1}}
\end{eqnarray}

An algebraic representation of the different beam shifts as 2D vectors for every combination of polarizer and analyser setting in the `$+\bigcirc\times$' bases is given in Table \ref{tab:algebraic}.
The table parts naturally in four blocks: the eigenpolarization block (`$+/+$'), in which both polarizer and analyser are set to either TE or TM, two blocks in which either the polariser or the analyser is set to an eigenpolarization (`$+/\bigcirc \times$') or (`$+/\bigcirc \times$'), and one block which contains the combinations of circular and diagonal polarizations (`$\bigcirc\times/\bigcirc\times$').
Because of the change of handedness on reflection we have switched left and right circular polarisation as analyser settings in the last block, which makes the whole table symmetric.
Symmetries along the rows of the table are therefore reflected in symmetries along the columns and it is often convenient to discuss only one. 

\begin{table}
\caption{\label{tab:algebraic}Table showing in an algebraic form the resulting complex shift vectors for the total shift in Eq.~(\ref{eq:finaltotaleq}) and and shift of a scalar component Eq.~(\ref{eq:compweak}) for different input polarizations $\boldsymbol{E}$ and in the case of the component shifts also different analyser polarizations $\boldsymbol{F}$. The formal\ae~for the expressions $d_p, d_s$ as well as $p, q, r, u, v, a, c, e, g, f, h$ are given in Table \ref{tab:shifts}.}

\begin{indented}

\item[]\begin{tabular}{cl!{\setlength{\arrayrulewidth}{1pt}\color{black}\vline}cc!{\color{black}\vline}cccc}
\br
\multicolumn{2}{l}{} & \multicolumn{6}{c}{Polarizer setting $\boldsymbol{E}$} \\
\\[-1ex]
\multicolumn{2}{l!{\setlength{\arrayrulewidth}{1pt}\color{black}\vline}}{Analyser} & \multicolumn{1}{>{\columncolor{red}[1pt]}c}{$\boldsymbol{\leftrightarrow}$} & \multicolumn{1}{>{\columncolor{mycyan}[1pt]}c}{$\boldsymbol{\updownarrow}$} & \multicolumn{1}{>{\columncolor{myyellow}[1pt]}c}{$\boldsymbol{\circlearrowleft}$} & \multicolumn{1}{>{\columncolor{blue}[1pt]}c}{\color{white}$\boldsymbol{\circlearrowright}$} & \multicolumn{1}{>{\columncolor{green}[1pt]}c}{$\boldsymbol{\neswarrow}$} & \multicolumn{1}{>{\columncolor{mymagenta}[1pt]}c}{$\boldsymbol{\nwsearrow}$} \\
\multicolumn{2}{l!{\setlength{\arrayrulewidth}{1pt}\color{black}\vline}}{setting $\boldsymbol{F}$} & TM & TE & RC & LC & 45$^\circ$ & 135$^\circ$ \\
\br
\multicolumn{2}{l!{\setlength{\arrayrulewidth}{1pt}\color{black}\vline}}{no analyser} & $\binom{d_p}{0}$ & $\binom{d_s}{0}$ & $\binom{p}{q}$ & $\binom{p}{-q}$ & $\binom{p}{r}$ & $\binom{p}{-r}$ \\ 
\br 
\multicolumn{1}{>{\columncolor{red}[1pt]}c}{$\boldsymbol{\leftrightarrow}$} & TM & $\binom{d_p}{0}$ & $\binom{0}{0}$ & $\binom{d_p}{\mathrm{i} u}$ & $\binom{d_p}{-\mathrm{i} u}$ & $\binom{d_p}{u}$ & $\binom{d_p}{-u}$ \\[-2ex]
\\
\multicolumn{1}{>{\columncolor{mycyan}[1pt]}c}{$\boldsymbol{\updownarrow}$} & TE & $\binom{0}{0}$ & $\binom{d_s}{0}$ & $\binom{d_s}{\mathrm{i}v}$ & $\binom{d_s}{-\mathrm{i}v}$ & $\binom{d_s}{-v}$ & $\binom{d_s}{v}$ \\
\mr
\multicolumn{1}{>{\columncolor{blue}[1pt]}c}{\color{white}$\boldsymbol{\circlearrowright}$} & LC & $\binom{d_p}{\mathrm{i}u}$ & $\binom{d_s}{\mathrm{i}v}$ & $\binom{a}{\mathrm{i}b}$ & $\binom{c}{0}$ & $\binom{e}{f}$ & $\binom{g}{-h}$ \\[-2ex]
\\
\multicolumn{1}{>{\columncolor{myyellow}[1pt]}c}{$\boldsymbol{\circlearrowleft}$} & RC & $\binom{d_p}{-\mathrm{i}u}$ & $\binom{d_s}{-\mathrm{i}v}$ & $\binom{c}{0}$ & $\binom{a}{-\mathrm{i}b}$ & $\binom{g}{h}$ & $\binom{e}{-f}$ \\[-2ex]
\\
\multicolumn{1}{>{\columncolor{green}[1pt]}c}{$\boldsymbol{\neswarrow}$} & 45$^\circ$ & $\binom{d_p}{u}$ & $\binom{d_s}{-v}$ & $\binom{e}{f}$ & $\binom{g}{h}$ & $\binom{c}{-d_s}$ & $\binom{a}{0}$ \\[-2ex]
\\
\multicolumn{1}{>{\columncolor{mymagenta}[1pt]}c}{$\boldsymbol{\nwsearrow}$} & 135$^\circ$ & $\binom{d_p}{-u}$ & $\binom{d_s}{v}$ & $\binom{g}{-h}$ & $\binom{e}{-f}$ & $\binom{a}{0}$ & $\binom{c}{d_s}$ \\
\br
\end{tabular}
\end{indented}
\end{table}

The table follows the notation in this paper in that the first component of each vector refers to the shift in the plane of incidence (longitudinal) and the second component to the shift out of the plane of incidence (transverse). 
As before the real part of each component determines the angular shift, while the imaginary part determines the spatial shift and this is why we may expect to find symmetries in the spatial shifts involving diagonal polarisation also mirrored in the angular shifts for the circular polarization.
For comparison the shift of the total beam, without filtering out a scalar component by the use of an analyser, but for different incident polarizations, is shown in the top row.

\begin{table}
\caption{\label{tab:shifts}Table giving the formal\ae~for the symbols in Table \ref{tab:algebraic}.
The formula is valid for both total internal and partial reflection, but the nature of the shift (angular, spatial or both) depends on the whether the formula yields a real, imaginary or generally complex number, and this is different for total internal and partial reflection.}
\begin{indented}
\item[]\begin{tabular}{ccclcl}
\br
Symbol	& Formula		& \multicolumn{2}{c}{Total internal reflection}	& \multicolumn{2}{c}{Partial reflection}	\\
\br
$d_p$	& $\frac{\overline{r}_p'}{\overline{r}_p}$	& 	& imaginary	&	& real	\\
$d_s$	& $\frac{\overline{r}_s'}{\overline{r}_s}$	&	& imaginary	& 	& real	\\
$p$		& $\frac{\overline{r}_p' \overline{r}_p^\ast + \overline{r}_s' \overline{r}_s^\ast}{|\overline{r}_p|^2 + |\overline{r}_s|^2}	$	&	&  imaginary	&	&  real	\\
$q$		& $\mathrm{i} \cot \theta_0 \frac{(\overline{r}_p + \overline{r}_s)(\overline{r}_p^\ast + \overline{r}_s^\ast)}{(|\overline{r}_p|^2 + |\overline{r}_s|^2)}$	&	&  imaginary	& 	& imaginary	\\
$r$		& $\cot \theta_0 \frac{(\overline{r}_p + \overline{r}_s)(\overline{r}_p^\ast - \overline{r}_s^\ast)}{|\overline{r}_p|^2 + |\overline{r}_s|^2}$	&	&  imaginary	&	&	 real	\\
$u$		& $\cot \theta_0 \frac{\overline{r}_p+\overline{r}_s}{\overline{r}_p}$	& $v^\ast$		& complex	& 	& real	\\
$v$		& $\cot \theta_0 \frac{\overline{r}_p+\overline{r}_s}{\overline{r}_s}$	& $u^\ast$	& complex	& 	& real	\\
$a$		& $\frac{\overline{r}_p' + \overline{r}_s'}{\overline{r}_p + \overline{r}_s}$	& $\mathrm{Im} \, a = -\mathrm{i} p$	& complex	& & real	\\
$b$		& $2\cos \theta_0$	&	&  real	&	&  real	\\
$c$		& $\frac{\overline{r}_p' - \overline{r}_s'}{\overline{r}_p - \overline{r}_s}$	& $\mathrm{Im} \, c = -\mathrm{i} p$	& complex	& & real	\\
$e$		& $\frac{\overline{r}_p' + \mathrm{i} \overline{r}_s'}{\overline{r}_p + \mathrm{i} \overline{r}_s}$	& $\mathrm{Im} \, e = -\mathrm{i} p$	& complex	& $g^\ast$	& complex	\\
$f$		& $(1 + \mathrm{i}) \cot \theta_0 \frac{\overline{r}_p + \overline{r}_s}{\overline{r}_p - \mathrm{i} \overline{r}_s}$	&	& imaginary	& $h^\ast$	& complex	\\
$g$		& $\frac{\overline{r}_p' - \mathrm{i} \overline{r}_s'}{\overline{r}_p - \mathrm{i} \overline{r}_s}$	& $\mathrm{Im} \, g = -\mathrm{i} p$	& complex	& $e^\ast$	& complex	\\
$h$		& $(1 - \mathrm{i}) \cot \theta_0 \frac{\overline{r}_p + \overline{r}_s}{\overline{r}_p + \mathrm{i} \overline{r}_s}$	&	& imaginary	& $f^\ast$	& complex	\\
\br
\end{tabular}
\end{indented}
\end{table}

The entries in Table \ref{tab:algebraic} are derived solely from the linear algebra in section \ref{sec:formulae} and are valid for both total and partial reflection.
The most simple block to explain is the eigenpolarization block where the only shift of order $\delta$ are the known spatial or angular Goos-H\"anchen shifts for TM (p) or TE (s) polarization.
In the `$+/\bigcirc \times$' block (and therefore also in the `$+/\bigcirc \times$' block) it is interesting to note that all the longitudinal shifts are equal to the general Goos-H\"anchen shift $d_p$ or $d_s$ and that the transverse shift for all entries in this block are given by the same complex valued function $u$, the real and imaginary part of with contribute to the spatial and angular shifts of the circular and diagonal parts of this block respectively.
More importantly for both circular and diagonal polarizations the transverse shifts are equal and opposite between the two orthogonal polarizations.

The transverse part of the `$+/\bigcirc$' block contains the spin-Hall effect of light \cite{HostenKwiat:SCI:2008}, but from the symmetry of Table \ref{tab:algebraic} follows that there is an analogous but reversed effects, for which the two incident circular polarizations experience shifts of equal magnitude but opposite sign when postselected by the same eigenpolarization.
One noteworthy feature in the `$\bigcirc\times/\bigcirc\times$' block is the transverse shift for the `$\times/\times$' polarisation which is given by the longitudinal shift $d_s$. In fact this is not solely a consequence of the algebraic approach, but depends on the form of the Fresnel reflection coefficients.

Indeed more symmetries can be discovered if the functional form of the Fresnel coefficients and their derivatives are substituted for the algebraic expressions, but still without specifying if the reflection is total or partial.
Taking this into account, there is a material independent shift of a component in the `$\bigcirc/\bigcirc$' block, because $b$ is a simple geometric factor $b=2\cos \theta_0$.
As this is real valued, the transverse shift in this block is purely spatial and its sign is given by the handedness of the incident polarization. Despite the fact that the other circular polarization has no transverse shift, the transverse shift for the total beam is not material independent, because the weightings in Eq. (\ref{eq:finaltotaleq}) depend on the Fresnel coefficients.

In total internal reflection $\overline{r}_p$ and $\overline{r}_s$ are unimodular (i.e.~$|\overline{r}_p|^2=|\overline{r}_s|^2=1$), and the ratios $\overline{r}_p'/\overline{r}_p$ and $\overline{r}_s'/\overline{r}_s$ are therefore purely imaginary. 
In this case $d_p$ and $d_s$ denote the purely spatial shift first discovered by Goos and H\"anchen.
Moreover, the function $p = (\overline{r}_p^\ast \overline{r}_p' + \overline{r}_s^\ast \overline{r}_s')/2$ in the row without analyser is also purely imaginary and there is no longitudinal angular shift for circular or diagonal polarisation either.
Similarly, $q = \mathrm{i}(\overline{r}_p + \overline{r}_s)(\overline{r}_p^\ast + \overline{r}_s^\ast)\cot \theta_0/2$ and $r = (\overline{r}_p + \overline{r}_s)(\overline{r}_p^\ast - \overline{r}_s^\ast)\cot \theta_0/2$ are purely imaginary and this means that there is no angular shift in the transverse direction either. 
This is a known result as the angular shift is related to the derivatives of the moduli of the reflection coefficients which in total internal reflection are zero \cite{Merano+:NatPhot3:2009}.
In particular for incident circular polarisation the presence of a transverse spatial shift for the total beam is well known as the Imbert-Fedorov effect \cite{Fedorov:DAN105:1955,Imbert:PRD5:1972}.

In total internal reflection, $u = v^\ast = (\overline{r}_p+\overline{r}_s)\cot \theta_0/\overline{r}_p$ which is generally complex, and the scalar components can thus experience angular shifts while the total beam does not.
Within the `$+/\bigcirc\times$' block incident circular polarisation thus gives rise to both transverse spatial and transverse angular shifts, but the angular shifts for TM and TE postselection average to zero because the total beam shows no transverse angular shift. 
Similarly, $u$ and $-v$ have equal but opposite real parts which account for equal and opposite transverse angular shifts between the TM and TE scalar components for diagonal incident polarization. 
In the `$\bigcirc\times/\bigcirc\times$' block, $a,c,e,g$ all have the same imaginary part $-\mathrm{i} p$, which means that all the longitudinal spatial shifts in this block are identical, which of course is reflected by the fact that the purely imaginary function $p$ determines the longitudinal spatial shift for the total beam.

In partial reflection from a dielectric interface the Fresnel coefficients are purely real and therefore $d_p$ and $d_s$ are as well, and represent only longitudinal angular shifts.
Also, $p$ is purely real in this case, and there is no longitudinal spatial shift of the total beam for any incident polarization. 
However, $q$ is purely imaginary and the total beam therefore undergoes not only a longitudinal angular shift, but also a transverse spatial shift, which is of equal magnitude but opposite sign for incident left and right circular polarization.
On the other hand $r$ is purely real and the total beam for incident diagonal polarization experiences a general transverse and longitudinal angular shift but no spatial shift.
In the `$+/\bigcirc\times$' block there is no longer a direct relation between $u$ and $v$ for partial reflection.
Both functions are real and the TM and TE components of both circular and diagonal incident polarization experience a different transverse shift.
In the `$\bigcirc\times/\bigcirc\times$' block $a$ and $c$ are real, while $e = g^\ast$ and generally complex. 
The presence of $e$ and $g$ in the `$\bigcirc/\times$' block is the cause for the `in plane spin separation of light' as discussed in \cite{Qin+:OE19:2011}.
Of course, because of the symmetry of the table a corresponding effect can be expected for a separation of diagonal polarisation components upon reflection of a circular polarised light beam and such an effect is evident in the presence of the same shifts in the `$\times/\bigcirc'$ block.

\begin{figure}
\begin{center}
\includegraphics[width=\textwidth]{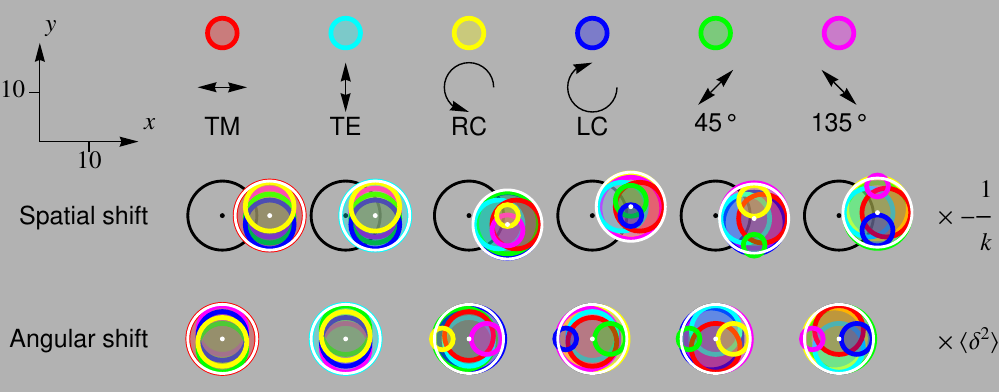}
\caption{\label{fig:comptir}
Table of the scalar component shifts in total internal reflection (n=2/3, $\theta_0 = \pi/4$). 
For each of the six incident polarization in the top row, the table shows the spatial and angular shifts for six corresponding polarization analyser settings, plotted on top of each other.
The size of the rings is proportional to the reflected intensity of this component, normalised to the total reflected intensity.
The shift is given in dimensionless unit and has to be multiplied by the appropriate measure ($-1/k$ and $\langle \delta^2 \rangle$) to give a physical quantity. The direction $x$ and $y$ are beam coordinates and refer to shifts with the plane (Goos-H\"anchen) and orthogonal to it (Imbert-Fedorov).
The size of the circles is not indicative of the scaling between the shift and the spot size of the optical beam. 
The black dot and circle \raisebox{-0.3ex}[0.1ex][0.1ex]{\includegraphics[height=2ex]{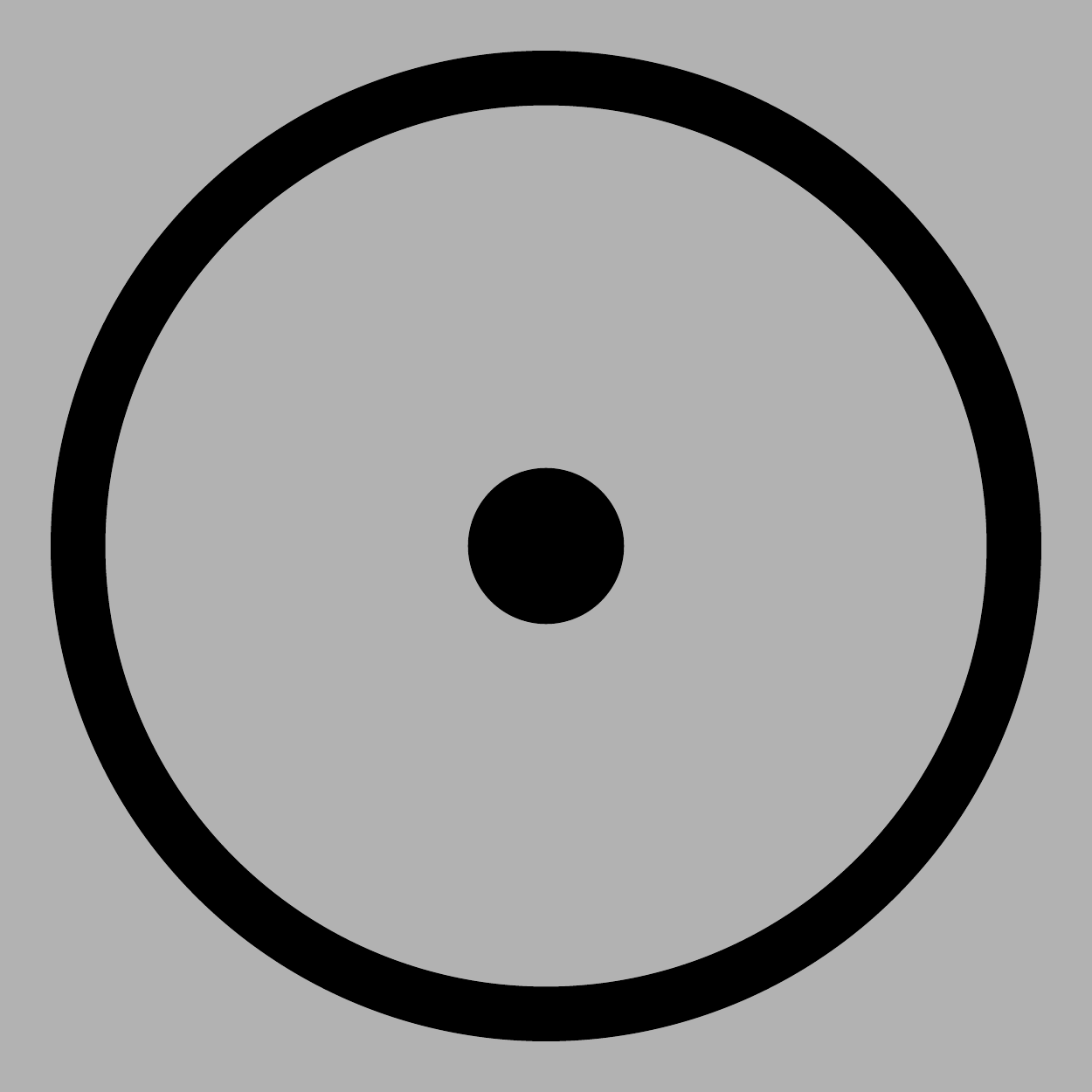}} indicates the position of the unshifted virtual reflected beam.
The white dot and outline circle \raisebox{-0.3ex}[0.1ex][0.1ex]{\includegraphics[height=2ex]{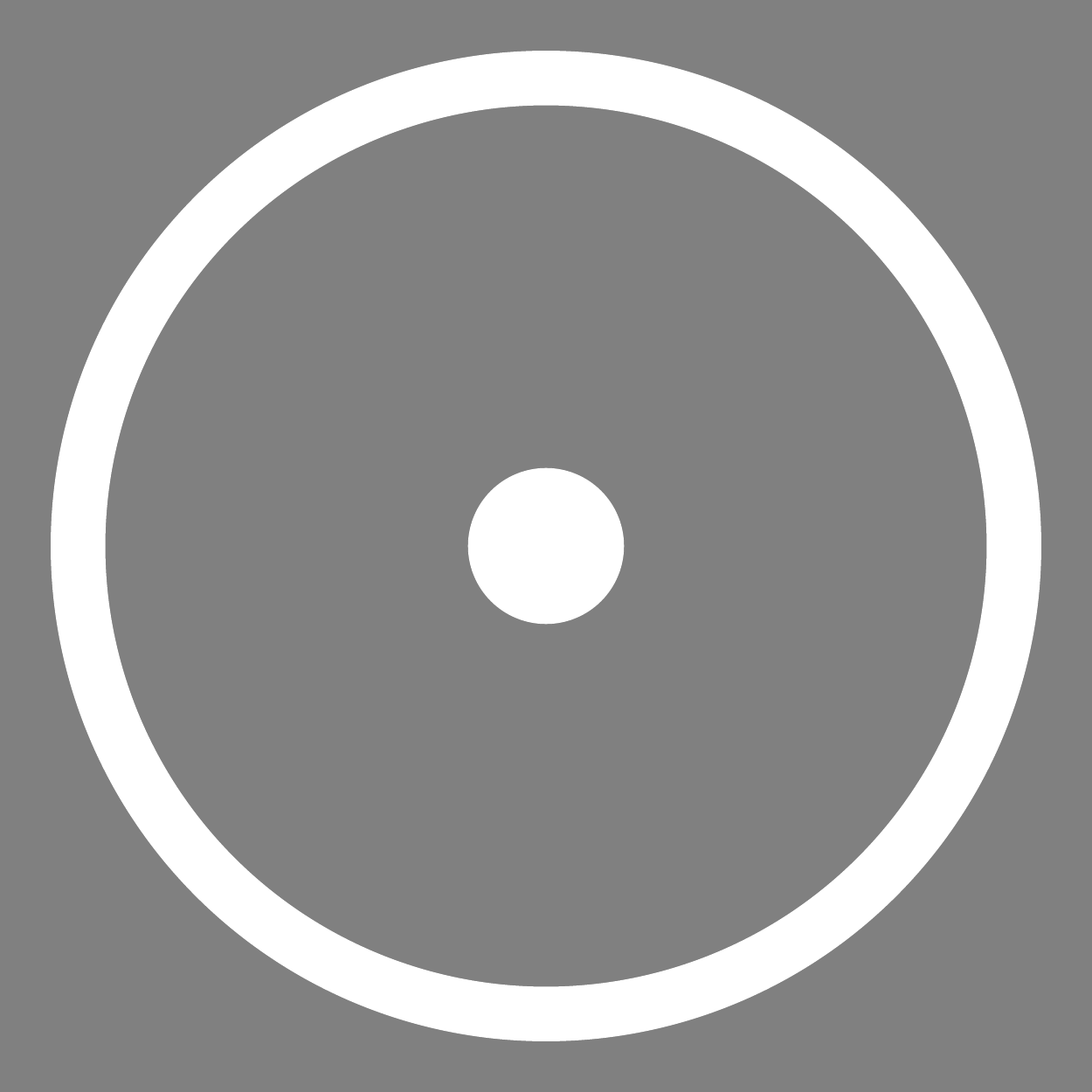}} marks the position and intensity of the total reflected beam, without filtering out a scalar component using a polarisation analyser.
   }
\end{center}
\end{figure}

\begin{figure}
\begin{center}
\includegraphics[width=\textwidth]{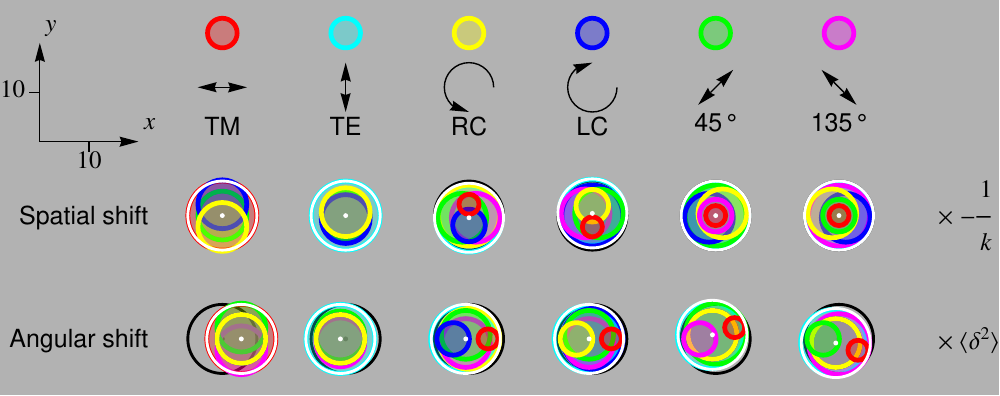}
\caption{\label{fig:comppr}Table of the scalar component shifts in external reflection (n=3/2, $\theta_0 = \pi/4$).
The table is constructed analogous to Fig. (\ref{fig:comptir}).
Because in partial reflection the Fresnel coefficients re purely real, any imaginary units are introduced by the circular polarization which explains why spatial shifts only occur when either the polarizer or the analyser selects circular polarization.} 
\end{center}
\end{figure}

A graphical representation of the shifts contained in this table is shown in Fig. \ref{fig:comptir}  for total internal reflection and in Fig.~\ref{fig:comppr} for partial reflection.
For both cases each row of Table \ref{tab:algebraic} is compressed into two plots of overlapping circles, representing the spatial and angular component shifts for a given incident polarization, refractive index ($n = 2/3$ for Fig. \ref{fig:comptir} and $n = 3/2$ for  Fig. \ref{fig:comppr}) and incidence angle of $\theta=\pi/4$ radians.
The scalar component for each analyser polarisation follows the same colour coding as the incident polarisation and also as in Table \ref{tab:algebraic}.
The position of the virtual reflected beam is indicated by a black circle and dot \raisebox{-0.3ex}[0.1ex][0.1ex]{\includegraphics[height=2ex]{VirtualCentre.pdf}}, whereas the shift of the total beam is indicated by a white circle and dot \raisebox{-0.3ex}[0.1ex][0.1ex]{\includegraphics[height=2ex]{TotalCentre.pdf}}.
The shifts are given as dimensionless quantities and have to be multiplied by $-1/k$ for the spatial shift and $\langle \delta^2 \rangle$ for the angular shift.

In addition to just quantifying the information given in Table \ref{tab:algebraic} these figures also show the intensity of the reflected scalar encoded in the area of the each circle.
The intensities are normalised with respect to the total reflected beam and should not be seen as indicative for the beam waist, which is usually so large that any spatial shifts are fully contained within its radius.
Being able to read out the intensity of a scalar component helps to understanding how shifts of different magnitudes between an orthogonal pair can average to zero for the total shift.
An example is the angular shift for circular polarisation in total internal reflection in Fig. \ref{fig:comptir}, where the component with identical incident and analyser polarisations experiences a larger shifts than for cross-polarization, and yet the total shift is zero because of the different intensities in each component.
This behaviour is fully in line with our understanding of beam shifts as realisations of weak values, where a lower intensity can go along with a larger value of the shift.


\section{Summary}\label{sec:disc}

We have presented here a self-contained approach to optical beams shift phenomena which results in a simple linear algebra of $2 \times 2$ matrices, in which the connection between optical beams shifts and weak values is readily established.
Moreover, on choosing three mutually incompatible bases for the incident and analyzer polarization ,we have been able to study the shifts of the centre of mass and the separate polarization components algebraically in a systemic manner, which resulted in Table \ref{tab:algebraic}.
Using this table we have highlighted the existent of symmetries between input and output polarizations, which means that well known effects, such as the spin-Hall effect of light \cite{HostenKwiat:SCI:2008} and the `in plane spin separation of light' \cite{Qin+:OE19:2011} are accompanied by analogous but reverse effects.
Further, we have found a material independent spatial, transverse shift, which can be observed for circular cross-polarization. 
As optical beams shifts can be used to gain information about the reflecting interface, the presence of a `zero', which does not depend on the material is naturally very important.
  
The connection of beam shifts and weak values is not new and has been pointed out before by Hosten and Kiwat \cite{HostenKwiat:SCI:2008} and phrased in a purely classical language by Aiello and Woerdman \cite{AielloWoerdman:OL33:2008}, but what hitherto has not been fully appreciated is, how exhaustive the interpretation of optical beam shifts phenomenon as weak values really is.
The interpretation of the angular shift as imaginary part of the weak value in section \ref{sec:formulae} makes it obvious that the natural measure for the angular shift is the second moment of the spectrum.
Using this as unit explains the difference in the angular Goos-H\"anchen (GH) shift between 3D vector beams \cite{Merano+:NatPhot3:2009}  and 2D `sheet' beams \cite{Merano+:OL21:2010} and also accounts for the recently observed radial mode dependence for the angular GH and Imbert-Fedorov (IF) shifts for general Laguerre-Gaussian modes \cite{Hermosa+:arXiv:2011}.
The vortex-induced shifts discussed in this paper can also be explained as weak values, but the full extent of the analogies between quantum weak measurements and optical beam shifts deserves a separate exposition elsewhere as it can discussed without reference to a given optical beam and material interface \cite{DennisGoette:NJP:2012}.


\section*{Acknowledgements}

We would like to thank Han Woerdman, Konstantin Bliokh, Andrea Aiello, Wolfgang L\"offler and Martina Hentschel for insightful discussions and encouragement. JBG and MRD acknowledge financial support from the Royal Society. A considerable part of this work was conducted while JBG was a Royal Society Newton Fellow at the University of Bristol.


\section*{References}


\begin{thebibliography}{10}

\bibitem{GoosHaenchen:AndP436:1947}
Goos F and H{\"a}nchen H
\newblock 1947
\newblock Ein neuer und fundamentaler Versuch zur Totalreflexion
\newblock {\em Annalen der Physik} {\bf 436} 333--46


\bibitem{Imbert:PRD5:1972}
Imbert C
\newblock 1972
\newblock Calculation and experimental proof of the transverse shift induced by total internal reflection of a circularly polarized light beam.
\newblock {\em Phys Rev D} {\bf 5} 787--96


\bibitem{Fedorov:DAN105:1955}
Fedorov F I
\newblock 1955
\newblock On the theory of total internal reflection
\newblock {\em Doklady Akademii Nauk SSSR} {\bf 105} 465--69


\bibitem{LibermanZeldovich:PRA46:1992}
Liberman V S and Zel'dovich B Y
\newblock 1992
\newblock Spin-orbit interaction of a photon in an inhomogeneous medium
\newblock {\em Phys Rev A} {\bf 46} 5199--207


\bibitem{BliokhBliokh:PRL96:2006}
Bliokh K Y and Bliokh Y P
\newblock 2006
\newblock Conservation of angular momentum, transverse shift, and spin Hall effect in reflection and refraction of an electromagnetic wave packet
\newblock {\em Phys Rev Lett} {\bf 96} 073903


\bibitem{ChanTamir:OL10:1985}
Chan C C and Tamir T
\newblock 1985
\newblock Angular shift of a Gaussian beam reflected near the Brewster angle
\newblock {\em Opt Lett} {\bf 10} 378--80


\bibitem{Merano+:NatPhot3:2009}
Merano M, Aiello A, van Exter M P and Woerdman J P
\newblock 2009
\newblock Observing angular deviations in the specular reflection of a light beam
\newblock {\em Nature Photonics} {\bf 3} 337--40


\bibitem{AielloWoerdman:OL33:2008}
Aiello A and Woerdman J P
\newblock 2008
\newblock Role of beam propagation in Goos-H{\"a}nchen and Imbert-Fedorov shifts
\newblock {\em Opt Lett} {\bf 33} 1437--9


\bibitem{HostenKwiat:SCI:2008}
Hosten O and Kwiat P
\newblock 2008
\newblock Observation of the spin Hall effect of light via weak measurements
\newblock {\em Science} {\bf 319} 787--90


\bibitem{Qin+:OE19:2011}
Qin Y, Li Y, Feng X, Xiao Y-F, Yang H and Gong Q
\newblock 2011
\newblock Observation of the in-plane spin separation of light
\newblock {\em Opt Exp} {\bf 19} 9636--45


\bibitem{Aharonov+:PRL60:1988}
Aharonov Y, Albert D Z, and Vaidman L
\newblock 1988
\newblock How the result of a measurement of a component of the spin of a spin-1/2 particle can turn out to be 100
\newblock {\em Phys Rev Lett} {\bf 60} 1351--4


\bibitem{AharonovRohrlich:WVCH:2005}
Yakir Aharonov Y and Rohrlich D
\newblock 2005
\newblock {\em Quantum Paradoxes}
\newblock Wiley


\bibitem{Steinberg:PRA52:1995}
Steinberg A M
\newblock 1995
\newblock Conditional probabilities in quantum theory and the tunneling-time controversy
\newblock {\em Phys Rev A} {\bf 52} 32--42


\bibitem{Josza:PRA76:2007}
Jozsa R
\newblock 2007
\newblock Complex weak values in quantum measurement
\newblock {\em Phys Rev A} {\bf 76} 044103


\bibitem{DennisGoette:NJP:2012}
Dennis M R and Goette J B
\newblock 2012
\newblock The analogy between optical beam shifts and quantum weak measurements
\newblock {\em submitted}


\bibitem{YinHesselink:APL89:2006}
Yin X and Hesselink L
\newblock 2006
\newblock Goos-H{\"a}nchen shift surface plasmon resonance sensor
\newblock {\em Appl Phys Lett} {\bf 89} 261108


\bibitem{Rodrigues-Herrera+:PRL104:2010}
Rodriguez-Herrera O G, Lara D, Bliokh K Y, Ostrovskaya E A and Dainty C
\newblock 2010 
\newblock Optical Nanoprobing via Spin-Orbit Interaction of Light 
\newblock {\em Phys Rev Lett} {\bf 104} 253601


\bibitem{BornWolf:CUP:2003}
Born M and Wolf E
\newblock 1999
\newblock {\em Principles of Optics}
\newblock Cambridge University Press, 7th edition


\bibitem{AielloWoerdman:OL36:2011}
Aiello A and Woerdman J P
\newblock 2011
\newblock Goos-H{\"a}nchen and Imbert-Fedorov shifts of a nondiffracting Bessel beam
\newblock {\em Opt Lett} {\bf 36} 543--5


\bibitem{Onada+:PRL93:2004}
Onoda M, Murakami S and Nagaosa N
\newblock 2004
\newblock Hall effect of light.
\newblock {\em Phys Rev Lett} {\bf 93} 083901


\bibitem{BerryShukla:JPA43:2010}
Berry M V and Shukla P
\newblock 2010
\newblock Typical weak and superweak values
\newblock {\em J Phys A: Math Theor} {\bf 43} 354024


\bibitem{AielloWoerdman:arxiv:2007}
Aiello A and Woerdman J P
\newblock 2009
\newblock Theory of angular Goos-H{\"a}nchen shift near Brewster incidence
\newblock arXiv:0903.3730


\bibitem{Aiello+:OL34:2009a}
Aiello A, Merano M and Woerdman J P
\newblock 2009
\newblock Brewster cross polarization
\newblock {\em Opt Lett} {\bf 34} 1207--9


\bibitem{GoetteDennis:OL:2012}
G{\"o}tte J B and Dennis M R
\newblock Superweak values for optical beam shifts at pseudo-Brewster angles
\newblock {\em in preparation}


\bibitem{Aiello+:PRL103:2009}
Aiello A, Lindlein N, Marquardt C and Leuchs G
\newblock 2009
\newblock Transverse angular momentum and geometric spin Hall effect of light
\newblock {\em Phys Rev Lett} {\bf 103} 100401


\bibitem{Merano+:OL21:2010}
Merano M, Hermosa N, Aiello A and Woerdman J P
\newblock 2010
\newblock Demonstration of a quasi-scalar angular Goos-H{\"a}nchen effect
\newblock {\em Opt Lett} {\bf 35} 3562--4


\bibitem{Hermosa+:arXiv:2011}
Hermosa N, Aiello A and Woerdman J P
\newblock 2012
\newblock Radial mode dependence of optical beam shifts
\newblock {\em Opt Lett} {\bf 37} 1044--6



\end{thebibliography}

\end{document}